\documentstyle[12pt]{article}
\newlength{\extraspace}
\newlength{\extraspaces}
\newcommand{\be}{\begin{equation}
\addtolength{\abovedisplayskip}{\extraspaces}
\addtolength{\belowdisplayskip}{\extraspaces}
\addtolength{\abovedisplayshortskip}{\extraspace}
\addtolength{\belowdisplayshortskip}{\extraspace}}
\newcommand{\ee}{\end{equation}}

\newcommand{\ba}{\begin{eqnarray}
\addtolength{\abovedisplayskip}{\extraspaces}
\addtolength{\belowdisplayskip}{\extraspaces}
\addtolength{\abovedisplayshortskip}{\extraspace}
\addtolength{\belowdisplayshortskip}{\extraspace}}
\newcommand{\ea}{\end{eqnarray}}

\newcommand{\newsection}[1]{
\vspace{15mm}
\pagebreak[3]
\addtocounter{section}{1}
\setcounter{equation}{0}
\setcounter{subsection}{0}
\setcounter{footnote}{0}
\begin{flushleft}
{\large\bf \thesection. #1}
\end{flushleft}
\nopagebreak
\medskip
\nopagebreak}

\newcommand{\Tr}{{\rm Tr}}

\begin{document}

\addtolength{\baselineskip}{.8mm}

{\thispagestyle{empty}
\noindent \hspace{1cm}  \hfill HD--THEP--98--9 \hspace{1cm}\\
\mbox{}                 \hfill February 1998 \hspace{1cm}\\

\begin{center}\vspace*{1.0cm}
{\large\bf The topological susceptibility of QCD:} \\
{\large\bf from Minkowskian to Euclidean theory} \\
\vspace*{1.0cm}
{\large Enrico Meggiolaro} \\
\vspace*{0.5cm}{\normalsize
{Institut f\"ur Theoretische Physik, \\
Universit\"at Heidelberg, \\
Philosophenweg 16, \\ 
D--69120 Heidelberg, Germany.}} \\
\vspace*{2cm}{\large \bf Abstract}
\end{center}

\noindent
We show how the topological susceptibility in the Minkowskian theory of QCD 
is related to the corresponding quantity in the Euclidean theory, which is 
measured on the lattice. We discuss both the zero--temperature case 
($T = 0$) and the finite--temperature case ($T \ne 0$). It is shown that 
the two quantities are equal when $T = 0$, while the relation between them 
is much less trivial when $T \ne 0$. The possible existence of 
``Kogut--Susskind poles'' in the matrix elements of the topological charge 
density between states with equal four--momenta turns out to invalidate
the equality of these two quantities in a strict sense. However, an equality 
relation is recovered after one re--defines the Minkowskian topological 
susceptibility by using a proper infrared regularization.
\vfill\eject

\newsection{Introduction}

\noindent
Since the pioneering works of Witten and Veneziano in 1979 for the solution 
of the ``$U(1)$ problem'' \cite{Witten79,Veneziano79}, a relevant role in the
non--perturbative study of QCD has been played by the so--called 
``topological susceptibility'', defined as (in the theory at zero 
temperature, $T=0$)
\be
\chi \equiv -i\int d^4 x \langle 0 | {\rm T}Q(x)Q(0) | 0 \rangle ~,
\ee
where Q(x) is the topological charge density operator:
\be
Q(x) \equiv {g^2 \over 64\pi^2} \varepsilon^{\mu\nu\rho\sigma}
F^a_{\mu\nu} F^a_{\rho\sigma} ~.
\ee
The topological susceptibility $\chi$ enters into the anomalous Ward 
identities of QCD \cite{Veneziano79,Crewther79} and its determination is of 
great relevance for understanding the role of the $U(1)$ axial symmetry in 
the spectrum and in the phase structure of the theory, both at zero and at 
finite temperature.
The lattice is a unique tool to determine from first principles quantities 
like $\chi$, which have non--trivial dimensions in mass and therefore 
cannot be computed by perturbation theory.
The first lattice determination of $\chi$ is that of Ref. 
\cite{DiVecchia-et-al.81}, where it was evaluated for the pure--gauge 
theory. After that ``seminal'' work, a lot of groups have developed refined 
lattice techniques to obtain more realistic estimates of $\chi$ (see, for
example, Refs. \cite{DiGiacomo97,vanBaal97} and references therein).

Anyway, to be rigorous, what is normally evaluated on the lattice is not
$\chi$ but the corresponding quantity $\chi_E$ defined in the Euclidean 
theory.
In this paper we discuss how the topological susceptibility $\chi$ in the 
Minkowski theory is related to the corresponding quantity $\chi_E$ in the 
Euclidean theory. We consider both the zero--temperature case ($T=0$) and 
the finite--temperature case ($T \ne 0$).
In Sect. 2 we shall prove that the two quantities $\chi$ and $\chi_E$ are 
equal when $T=0$. The case $T \ne 0$ is much less trivial and is discussed 
in detail in Sect. 3.

The finite temperature topological susceptibility $\chi (\beta)$ (where
$\beta = 1/kT$, $k$ being the Boltzmann constant) is defined as
\be
\chi (\beta) \equiv -i\int d^4 x \langle {\rm T}Q(x)Q(0)\rangle_\beta ~,
\ee
where $\langle \ldots \rangle_\beta$ is the usual quantum 
thermal average over the Gibbs ensemble 
\cite{Matsubara55,Abrikosov65,Kapusta89}.
Now the problem to be solved is the following. In Eq. (1.3) we have a 
finite--temperature Green function at different times, namely 
$\langle {\rm T}Q(x)Q(0) \rangle_\beta$,
which is then integrated over the entire Minkowskian space--time.
In order to do Monte Carlo simulations on the lattice to measure a certain 
Green function, one needs to represent such a Green function by a 
path--integral expression in the Euclidean theory. This is well known for 
Green functions at $T=0$ (see Sect. 2) and also for finite--temperature 
($T \ne 0$) Green functions at equal times, i.e., of the kind 
$\langle O(\vec{x}, 0) O(0, 0) \rangle_\beta$.
However it is not so immediate to express a finite temperature Green function 
at different times, i.e., 
$\langle {\rm T}O(\vec{x}, t) O(\vec{0}, 0) \rangle_\beta$, 
in terms of a path--integral. 
As discussed in Sect. 3, the Euclidean topological susceptibility 
$\chi_E (\beta)$ can be expressed as a path--integral over an Euclidean 
four--space limited in the time direction (so that it has the topology of 
${\it R}^3 \otimes {\it T}^1$, where ${\it T}^1$ is a one--dimensional thorus).
However, the relation between $\chi(\beta)$ and $\chi_E(\beta)$ turns out 
to be much less trivial than in the $T=0$ case. 
In particular, it comes out that the possible existence of 
``Kogut--Susskind poles'' in the matrix elements of the topological charge 
density between states with equal four--momenta could invalidate the equality 
of these two quantities in a strict sense. Nevertheless, it will be shown 
in Sect. 3 that an equality relation is recovered after one re--defines the 
Minkowskian topological susceptibility by using a proper infrared 
regularization. Finally, in the Appendix we shall give a more accurate
discussion about the correct definition of the topological susceptibility
with the inclusion of the so--called ``contact term'' (see Refs.
\cite{Witten79,Crewther79}).

\newsection{The zero--temperature case ($T = 0$)}

\noindent
Let us start with the most simple case: the zero--temperature one
($T = 0$). The topological susceptibility at $T=0$ is defined as
\be
\chi \equiv -i\int d^4x \langle 0 | {\rm T} Q(x) Q(0) | 0 \rangle ~,
\ee
where $Q(x)$ is the operator for the topological charge density and
${\rm T}Q(\vec{x}, t) Q(\vec{0}, 0)$ is the usual time--ordered product:
\be
{\rm T}Q(\vec{x}, t) Q(\vec{0}, 0) = \theta (t) Q(\vec{x}, t)Q(\vec{0}, 0) +
\theta (-t) Q(\vec{0}, 0)Q(\vec{x}, t) ~.
\ee
[Actually, Eq. (2.1) is not the correct definition of the topological
susceptibility: a certain equal--time commutator term (also called
``contact term'') must be added to the right--hand side
\cite{Witten79,Crewther79}. A more careful discussion about the correct
formula including the contact term is given in the Appendix. As it will be
shown in the Appendix, all the results we find are not affected by the
inclusion of the contact term. This last, however, solves some questions
about positivity arising in connection with some of the formulas below.]

Remembering that, in the Heisenberg representation of quantum 
operators, $Q(\vec{x}, t) = e^{iHt} Q(\vec{x}, 0) e^{-iHt}$ ($H$ being the 
total Hamiltonian operator), we may insert between the two $Q$'s a complete 
set of eigenstates $\{|n\rangle\}$ which diagonalize the total 
four--momentum operator $P^\mu$ (i.e., $P^\mu |n\rangle = q^\mu_n 
|n\rangle$; remember that $[P^\mu,P^\nu] = 0$, with $P^0 \equiv H$.)
Then we may perform the integrations in the space coordinates as well as in 
the time coordinate, making use (in this last case) of the following relations:
\ba
\displaystyle\int_{0}^{+\infty} e^{i\alpha t} dt &=& {\rm P}
{ i \over \alpha} + \pi \delta (\alpha) ~, \nonumber \\
\displaystyle\int_{-\infty}^{0} e^{i\alpha t} dt &=& -{\rm P}
{ i \over \alpha} + \pi \delta (\alpha) ~.
\ea
(``P'' stands for ``principal part''.)
We thus find the following expression:
\be
\chi = -2(2\pi)^3 \displaystyle\sum_{n \ne 0}
{1 \over q_n^0} |\langle 0 | Q(0) | n \rangle|^2 
\delta^{(3)} (\vec{q}_n) ~.
\ee
In the derivation of this formula we have also made use of the fact that
$Q(x)$ is an hermitean operator and that its vacuum expectation value is 
zero for parity invariance ($\langle 0| Q(x) |0 \rangle = 0$ since $Q(x)$ 
is odd under parity transformations).
We have simbolically indicated a ``sum'' over a complete set of eigenstates
$|n\rangle$ with four--momenta $q^\mu_n$, but this clearly means that one has 
to sum over the discrete modes and integrate over the continuous distribution 
of eigenstates.

More generally, one can consider $\chi$ as the value at zero four--momentum
($p=0$) of the Fourier transform of the two--point Green function
$I(x) \equiv -i\langle 0 | {\rm T} Q(x) Q(0) | 0 \rangle$;
i.e., $\chi = \tilde{I}(p=0)$, where
\be
\tilde{I}(p) \equiv -i\int d^4x e^{ipx} \langle 0 | {\rm T} Q(x) Q(0) | 
0 \rangle ~.
\ee
This quantity can be written in an elegant way in terms of the so--called 
``spectral density'' $\rho (p)$ of $I(x)$:
\be
\rho (p) \equiv (2\pi)^3 \displaystyle\sum_n |\langle 0 | Q(0) | n \rangle|^2
\delta^{(4)} (p-q_n) ~.
\ee
Thanks to the fact that $\langle 0| Q(x) |0 \rangle = 0$, the contribution 
from the vacuum states ($n=0$) vanishes, that is:
\be
\rho_0 (p) \equiv (2\pi)^3 |\langle 0 | Q(0) | 0 \rangle|^2 \delta^{(4)} (p) 
= 0 ~,
\ee
and we are left with
\be
{\bar \rho} (p) \equiv (2\pi)^3 \displaystyle\sum_{n \ne 0} 
|\langle 0 | Q(0) | n \rangle|^2 \delta^{(4)} (p-q_n) ~.
\ee
For covariance reasons, it is clear that
\be
{\bar \rho} (p) = {\bar \sigma} (p^2) {\bar \theta} (p^0) ~,
\ee
where ${\bar \theta} (p^0) = 1$ for $p^0 > 0$ and ${\bar \theta} (p^0) = 0$ 
for $p^0 \le 0$. Because of the ``natural'' conditions which are assumed on
the spectrum of $P^\mu$, ${\bar \sigma} (p^2)$ vanishes for $p^2 < 0$ and
is real and positive semidefinite for $p^2 \ge 0$.
One can then write $\tilde{I}(p)$ in the so--called ``spectral 
representation'', firstly derived by K\"allen and Lehmann in Refs.
\cite{Kallen52,Lehmann54}:
\be
\tilde{I}(p) = \displaystyle\int_0^{+\infty} d(\mu^2)
{{\bar \sigma}(\mu^2) {\bar \theta}(\sqrt{\mu^2 + \vec{p}^2}) \over
p^2 - \mu^2 + i\varepsilon} ~.
\ee
Therefore, an alternative expression for $\chi$ is also derived:
\be
\chi = \tilde{I}(p=0) = -\displaystyle\int_0^{+\infty} d(\mu^2)
{\bar \sigma}(\mu^2) {\rm P} {1 \over \mu^2} ~.
\ee
Let us consider, now, the Euclidean theory. We may define 
$Q(\vec{x}, \tau) = e^{iH\tau} Q(\vec{x}, 0) e^{-iH\tau}$
for every complex $\tau$. In particular, we shall consider the following 
quantity integrated over the imaginary axis from $+i\infty$ to $-i\infty$:
\be
\tilde{G}(p) \equiv -\displaystyle\int_{-\infty}^{+\infty} d\tau
\displaystyle\int d^3\vec{x} e^{ip^0\tau - i\vec{p} \cdot \vec{x}}
\langle 0 | {\rm T} Q(\vec{x},-i\tau) Q(\vec{0},0) | 0 \rangle ~.
\ee
The following prescription for the T--ordered product of a bosonic field
$B$ in the imaginary domain is used:
\ba
{\rm T} B(\tau_1) B(\tau_2) &=& B(\tau_1) B(\tau_2), ~~~{\rm if}~i\tau_1 > 
i\tau_2 ~; \nonumber \\
{\rm T} B(\tau_1) B(\tau_2) &=& B(\tau_2) B(\tau_1), ~~~{\rm if}~i\tau_1 < 
i\tau_2 ~. 
\ea
In other words, $\theta (-i t) \equiv \theta (t)$, for every real $t$: 
this prescription is used in order to keep the T--ordering unchanged when
going from Minkowskian to Euclidean theory, 
$(x^0,\vec{x}) \rightarrow (-ix_{E4},\vec{x}_E)$. 
Proceeding as for the derivation of Eq. (2.4), we can derive the 
following expression for $\tilde{G}(p)$:
\be
\tilde{G}(p) = -2(2\pi)^3 \displaystyle\sum_{n \ne 0}
{q_n^0 \over (q_n^0)^2 + (p^0)^2} |\langle 0 | Q(0) | n \rangle|^2 
\delta^{(3)} (\vec{p} - \vec{q}_n) ~.
\ee
Therefore, at zero four--momentum ($p=0$) we find exactly the same 
expression reported in Eq. (2.4) for $\tilde{I}(p=0)$:
\be
\tilde{G}(p=0) = -2(2\pi)^3 \displaystyle\sum_{n \ne 0}
{1 \over q_n^0} |\langle 0 | Q(0) | n \rangle|^2 
\delta^{(3)} (\vec{q}_n) = \tilde{I}(p=0) ~.
\ee
It is easy to see that $\tilde{G}(p=0)$ is the topological 
susceptibility $\chi_E$ in the Euclidean theory. In fact:
\ba
\lefteqn{
\tilde{G}(p=0) =
-\displaystyle\int_{-\infty}^{+\infty} d\tau \displaystyle\int d^3\vec{x}
\langle 0 | {\rm T} Q(\vec{x},-i\tau) Q(\vec{0},0) | 0 \rangle }
\nonumber \\
& & = \displaystyle\int d^4 x_E \langle Q_E(x_E) Q_E(0) \rangle_E 
\equiv \chi_E ~,
\ea
where
\be
\langle Q_E(\vec{x},\tau) Q_E(\vec{0},0) \rangle_E \equiv 
-\langle 0 | {\rm T} Q(\vec{x},-i\tau) Q(\vec{0},0) | 0 \rangle
\ee
(``$E$'' stands for ``Euclidean'') is just the Green function in the 
Euclidean theory. [A more accurate discussion about Eqs. (2.16) and (2.17)
with the inclusion of the contact term is given in the Appendix.]
We remind that the correspondence relationships from 
Minkowskian to Euclidean theory are:
\ba
(x^0,\vec{x}) &\rightarrow& (-ix_{E4},\vec{x}_E) ~; \nonumber \\
A_0(x) &\rightarrow& iA_{E 4}(x_E) ~; \nonumber \\  
A_k(x) &\rightarrow& A_{E k}(x_E) ~~~ (k = 1,2,3) ~,
\ea
with $x_E = (\vec{x}_E, x_{E4})$. And therefore:
\be
Q(x) \rightarrow iQ_E(x_E) ~.
\ee
From Eqs. (2.16), (2.15) and (2.4) we finally derive that 
\be
\chi = \chi_E ~,
\ee
i.e., the topological susceptibility in the Minkowskian theory coincides 
with the topological susceptibility in the Euclidean theory. This last is 
the quantity which can be measured on the lattice
\cite{DiGiacomo97,vanBaal97}.

\newsection{The finite--temperature case ($T \ne 0$)}

\noindent
Let us address, now, the most difficult case: the finite--temperature case
($t \ne 0$). 
The finite temperature topological susceptibility $\chi (\beta)$ (where
$\beta = 1/kT$, $k$ being the Boltzmann constant) is defined as
\be
\chi (\beta) \equiv -i\int d^4 x \langle {\rm T}Q(x)Q(0)\rangle_\beta ~.
\ee
[Actually, as in the $T=0$ case, Eq. (3.1) is not quite correct and a 
certain contact term must be added to the right--hand side. We refer again
to the Appendix for a more accurate discussion about this point.]

The expectation value $\langle \ldots \rangle_\beta$ is the usual quantum 
thermal average over the Gibbs ensemble 
\cite{Matsubara55,Abrikosov65,Kapusta89}:
\be
\langle O \rangle_\beta \equiv
{\Tr [e^{-\beta H} O] \over Z(\beta)} ~,
\ee
where
\be
Z(\beta) \equiv \Tr [e^{-\beta H}]
\ee
is the partition function of the system. As before, $\chi(\beta)$ can be 
seen as the value at zero four--momentum ($p=0$) of the Fourier transform 
of the two--point Green function at finite temperature
$I_\beta (x) = -i\langle {\rm T} Q(x)Q(0) \rangle_\beta$; that is,
$\chi(\beta) = \tilde{I}_\beta (p=0)$, where:
\be
\tilde{I}_\beta (p) \equiv -i\displaystyle\int d^4 x e^{ipx} 
\langle {\rm T} Q(x)Q(0) \rangle_\beta ~.
\ee
One can derive an expression for $\tilde{I}_\beta (p)$ in terms of the
finite--temperature spectral density $\rho_\beta (p)$, defined as
\be
\rho_\beta (p) \equiv {(2\pi)^3 \over Z(\beta)} \displaystyle\sum_{m,n}
e^{-\beta q_m^0} |\langle m | Q(0) | n \rangle|^2
\delta^{(4)} (p+q_m-q_n) ~.
\ee
It is easy to see, from covariance arguments and from the explicit 
expression (3.5), that the spectral density $\rho_\beta(p)$ has the following 
properties:
\ba
\rho_\beta (p) &=& \rho_\beta (p^2,p^0) ~; \nonumber \\
\rho_\beta (p^2,-p^0) &=& e^{-\beta p^0} \rho_\beta (p^2,p^0) ~.
\ea
Analogously to the zero--temperature case, one can separate within 
$\rho_\beta(p)$ the contribution at $p^0=0$:
\be
\rho_\beta (p^2,p^0) = \rho_{\beta,0} (p^2,p^0) + 
{\bar \rho}_\beta (p^2,p^0) ~,
\ee
where:
\ba
\rho_{\beta,0} (p^2,p^0) &\equiv& {(2\pi)^3 \over Z(\beta)} 
\displaystyle\sum_{q_m^0 = q_n^0} e^{-\beta q_m^0} 
|\langle m | Q(0) | n \rangle|^2 \delta^{(4)} (p+q_m-q_n) ~; \nonumber \\
{\bar \rho}_\beta (p^2,p^0) &\equiv& {(2\pi)^3 \over Z(\beta)}
\displaystyle\sum_{q_m^0 \ne q_n^0} e^{-\beta q_m^0} 
|\langle m | Q(0) | n \rangle|^2 \delta^{(4)} (p+q_m-q_n) ~.
\ea
Therefore, by definition:
\ba
\rho_{\beta,0} (p^2,p^0) &=& 0, ~~~{\rm when}~~ p^0 \ne 0 ~; \nonumber \\
{\bar \rho}_\beta (p^2,p^0) &=& 0, ~~~{\rm when}~~ p^0 = 0 ~.
\ea
In terms of these two quantities, the following expression for
$\tilde{I}_\beta(p)$ can be derived:
\ba
\lefteqn{
\tilde{I}_\beta (p) = \displaystyle\int_0^{\infty} d(E^2)
{\bar \rho}_\beta (E^2-\vec{p}^2,E) (1-e^{-\beta E}) \times }
\nonumber \\
& & \times \left[ {1 \over p_0^2 - E^2 + i\varepsilon} 
-i{2\pi \over e^{\beta E} - 1} \delta(p_0^2-E^2) \right] 
-2\pi i \rho_{\beta,0} (p^2,p_0) ~.
\ea
Therefore, at zero four--momentum ($p=0$):
\be
\chi(\beta) = \tilde{I}_\beta (p=0) = -\displaystyle\int_0^{\infty} d(E^2)
{\bar \rho}_\beta (E^2,E) (1-e^{-\beta E}) {\rm P} {1 \over E^2}
-2\pi i \rho_{\beta,0} (0,0) ~.
\ee
In other words, separating the real part and the imaginary part:
\ba
\lefteqn{
{\rm Re} [\chi(\beta)] = {\rm Re} [\tilde{I}_\beta (p=0)] } \nonumber \\
& & = -\displaystyle\int_0^{\infty} d(E^2)
{\bar \rho}_\beta (E^2,E) (1-e^{-\beta E}) {\rm P} {1 \over E^2}
\nonumber \\
& & = {(2\pi)^3 \over Z(\beta)} \displaystyle\sum_{q_m^0 \ne q_n^0}
{e^{-\beta q_m^0} - e^{-\beta q_n^0} \over q_m^0 - q_n^0} 
|\langle m | Q(0) | n \rangle|^2 \delta^{(3)} (\vec{q}_m-\vec{q}_n) ~.
\ea
And also:
\ba
\lefteqn{
{\rm Im} [\chi(\beta)] =
{\rm Im} [\tilde{I}_\beta (p=0)] = -2\pi \rho_{\beta,0} (0,0) }
\nonumber \\
& & = -{(2\pi)^4 \over Z(\beta)} \displaystyle\sum_{m,n} e^{-\beta q_m^0} 
|\langle m | Q(0) | n \rangle|^2 \delta^{(4)} (q_m-q_n) ~.
\ea
In order to compare Eq. (3.11) with the corresponding quantity in the Euclidean 
theory, we shall proceed as in the previous Section and consider the 
following quantity, integrated over the imaginary axis from $0$ to
$-i\beta$:
\be
\tilde{G}_\beta (p) \equiv -\displaystyle\int_0^\beta d\tau
\displaystyle\int d^3 \vec{x} e^{ip^0\tau -i\vec{p} \cdot \vec{x}}
\langle {\rm T} Q(\vec{x},-i\tau) Q(\vec{0},0) \rangle_\beta ~,
\ee
where the prescription for the T--ordered product is given by Eq. (2.13). We 
remind that the correlation function $\langle B(t) B(0) \rangle_\beta$ is 
analytical for $-\beta < {\rm Im}(t) < 0$, so that the integral in Eq. 
(3.14) is taken over the analyticity domain of
$\langle Q(\vec{x},-i\tau) Q(\vec{0},0) \rangle_\beta$.
The following expression for $\tilde{G}_\beta (p)$ can be derived:
\be
\tilde{G}_\beta (p) = {(2\pi)^3 \over Z(\beta)} \displaystyle\sum_{m,n}
{e^{-\beta q_m^0} - e^{-\beta q_n^0 +i\beta p^0} \over q_m^0 - q_n^0 + ip^0} 
|\langle m | Q(0) | n \rangle|^2 \delta^{(3)} (\vec{p} + \vec{q}_m - \vec{q}_n)
~.
\ee
In particular, at $p=0$:
\ba
\lefteqn{
\tilde{G}_\beta (p=0) = } \nonumber \\
& & = {(2\pi)^3 \over Z(\beta)} \displaystyle\sum_{q_m^0 \ne q_n^0}
e^{-\beta q_m^0} \left[ {1 - e^{\beta (q_m^0 - q_n^0)}
\over q_m^0 - q_n^0} \right] |\langle m | Q(0) | n \rangle|^2 
\delta^{(3)} (\vec{q}_m - \vec{q}_n) \nonumber \\
& & -\beta {(2\pi)^3 \over Z(\beta)} \displaystyle\sum_{q_m^0 = q_n^0}
e^{-\beta q_m^0} |\langle m | Q(0) | n \rangle|^2 
\delta^{(3)} (\vec{q}_m - \vec{q}_n) ~.
\ea
Therefore, from Eqs. (3.16), (3.12) and (3.13):
\ba
\tilde{G}_\beta (p=0) &=& {\rm Re}[\chi(\beta)]
- \beta {(2\pi)^3 \over Z(\beta)} \displaystyle\sum_{q_m^0 = q_n^0}
e^{-\beta q_m^0} |\langle m | Q(0) | n \rangle|^2 
\delta^{(3)} (\vec{q}_m - \vec{q}_n) ~; \nonumber \\
{\rm Im}[\chi(\beta)] &=&
- {(2\pi)^4 \over Z(\beta)} \displaystyle\sum_{m,n}
e^{-\beta q_m^0} |\langle m | Q(0) | n \rangle|^2 
\delta^{(4)} (q_m - q_n) ~.
\ea
An alternative way to derive these equations starts from considering some 
basic properties of finite--temperature Green functions
\cite{Abrikosov65,Kapusta89}. In particular, one 
uses the fact that the correlation function $\langle B(t) B(0) \rangle_\beta$
is analytical for $-\beta < {\rm Im}(t) < 0$, while
$\langle B(0) B(t) \rangle_\beta$ is analytical for $0 < {\rm Im}(t) < \beta$. 
Moreover, one finds the following boundary conditions in the analyticity 
domains (if $B$ is a bosonic field):
\ba
\langle B(0) B(t) \rangle_\beta &=& \langle B(t-i\beta ) B(0) \rangle_\beta ~,
\nonumber \\
\langle B(t) B(0) \rangle_\beta &=& \langle B(0) B(t+i\beta ) \rangle_\beta ~,
\ea
valid for every real $t$.
By virtue of all these results, one may apply the Cauchy theorem when 
integrating $\langle Q(\vec{x},\tau ) Q(\vec{0}, 0) \rangle_\beta $ 
along the path represented in Fig. 1.
One thus finds the same result (3.17), using also the fact that:
\ba
\lefteqn{
\displaystyle\lim_{{\rm T} {\to} +\infty} 
\displaystyle\int_{{\rm T}}^{{\rm T}-i\beta} d\tau
\int d^3\vec{x} \langle Q(\vec{x}, \tau) Q(\vec{0}, 0) \rangle_\beta = } 
\nonumber \\
& & = -i\beta {(2\pi)^3 \over Z(\beta)}
\displaystyle\sum_{q_m^0 = q_n^0} e^{-\beta q_m^0} 
\vert \langle m\vert Q(\vec{0}, 0) \vert n \rangle \vert ^2  
\delta^{(3)} (\vec{q}_m - \vec{q}_n) ~.
\ea
Now let us observe that, by virtue of the correspondence relationships 
(2.18) and (2.19) from Minkowskian to Euclidean theory:
\be
-\langle {\rm T} Q(\vec{x}, -i\tau) Q(\vec{0}, 0) \rangle_\beta \equiv
\langle Q_E(\vec{x}, \tau) Q_E(\vec{0}, 0) \rangle_{E,\beta} \equiv 
M^{(E)}_\beta (\vec{x}, \tau) ~.
\ee
This is just the Matsubara finite--temperature Green function 
\cite{Matsubara55}, defined in the Euclidean theory, 
with $0 \le \tau \le \beta$. 
Therefore $\tilde{G}_\beta (p=0)$ is the Euclidean topological 
susceptibility at finite temperature, which we shall indicate with
$\chi_E(\beta)$:
\ba
\lefteqn{
\tilde{G}_\beta (p=0) 
= -\displaystyle\int_0^\beta d\tau \displaystyle\int d^3 \vec{x} 
\langle {\rm T} Q(\vec{x},-i\tau) Q(\vec{0},0) \rangle_\beta }
\nonumber \\
& & = \displaystyle\int_0^\beta d\tau \displaystyle\int d^3 \vec{x} 
\langle Q_E (\vec{x},\tau) Q_E (\vec{0},0) \rangle_{E,\beta}
\equiv \chi_E (\beta) ~.
\ea
[A more accurate discussion about Eqs. (3.20) and (3.21) with the inclusion
of the contact term is given in the Appendix.]
This is the quantity which can be measured on the lattice, thanks to the 
fact that a Matsubara Green function, such as 
$M^{(E)}_\beta (\vec{x}, \tau)$, admits a representation in a 
path--integral form as follows (see, for example, Ref. \cite{Bernard74}):
\be
M^{(E)}_\beta (\vec{x}, \tau) = {\int_{periodic}[dA_E] 
\int_{anti-p.}[d\psi_E] [d{\bar \psi}_E] Q_E(\vec{x}, \tau) Q_E(\vec{0}, 0) 
e^{-\int_{0}^{\beta}d\tau' \int d^3\vec{x}' L_E} \over Z(\beta) } ~,
\ee
where $L_E$ is the Euclidean lagrangian density of QCD. The partition 
function itself may be written as a path--integral over all the Euclidean 
gauge--field configurations which are periodic, with period $\beta$, in the 
Euclidean time, and over all the fermion configurations which are 
anti--periodic, with period $\beta$, in the Euclidean time \cite{Bernard74}:
\be
Z(\beta) \equiv \Tr[e^{-\beta H}] = 
\int_{periodic}[dA_E] \int_{anti-p.}[d\psi_E] [d{\bar \psi}_E] 
e^{-\int_{0}^{\beta}d\tau \int d^3\vec{x} L_E} ~.
\ee
Finally we have the following path--integral expression for the Euclidean 
topological susceptibility at finite temperature $\chi_E (\beta)$:
\be
\chi_E (\beta) = \displaystyle\int_{0}^{\beta} d\tau \int d^3 \vec{x}
{\int_{periodic}[dA_E] \int_{anti-p.}[d\psi_E] [d{\bar \psi}_E] 
Q_E(\vec{x}, \tau) Q_E(\vec{0}, 0)
e^{-\int_{0}^{\beta}d\tau' \int d^3 \vec{x}' L_E} \over
\int_{periodic}[dA_E] \int_{anti-p.}[d\psi_E] [d{\bar \psi}_E] 
e^{-\int_{0}^{\beta}d\tau' \int d^3 \vec{x}' L_E} } 
~.
\ee
This expression indicates that, in the lattice formulation of the theory, 
one must use an asymmetrical lattice of the form $N_\sigma^3 \times N_\tau$, 
with a number of time sites $N_\tau$ (much) smaller than the number of 
sites $N_\sigma$ along a space coordinate axis.
In this way, $a \cdot N_\tau = \beta$ and $(a \cdot N_\sigma)^3 = V \to
\infty$, where $a$ is the lattice spacing and $V$ is the space volume, 
which must be sent to infinity in the thermodynamical limit
(see Ref. \cite{Rothe92} and references therein). 

Let us compare, now, $\chi_E(\beta)$ with the ``physical'' quantity 
$\chi(\beta)$ in the Minkowskian theory.
At first sight, the two quantities $\chi(\beta) = \tilde{I}_\beta
(p=0)$, given by Eq. (3.11), and $\chi_E(\beta) = \tilde{G}_\beta
(p=0)$, given by Eq. (3.16), are not equal.
From Eq. (3.17) it is clear that a crucial role, in the relation between
$\chi(\beta)$ and $\chi_E(\beta)$, is played by those terms which contain 
the expectation values $\langle m| Q(0) |n \rangle$ between states with 
equal four--momenta, $q_m = q_n$.
It is known, however, that $Q(x)$ is the four--divergence of a four--current, 
i.e., $Q(x) = \partial^\mu K_\mu (x)$, with
\be
K_\mu \equiv {g^2 \over 16\pi^2} \varepsilon _{\mu \alpha \beta \gamma} A^
\alpha _a (\partial^\beta A^\gamma _a -{1 \over 3} g f_{abc} 
A^\beta _b A^\gamma _c) ~,
\ee
where $\mu,\alpha,\beta,\gamma$ are Lorentz indices, $a,b,c$ are colour 
indices and $f_{abc}$ are the structure constants of the colour group.
Therefore we deduce that:
\be
\langle m\vert Q(0) \vert n \rangle = i(q_m - q_n)^\mu
\langle m\vert K_\mu (0) \vert n \rangle ~.
\ee
Normally, if $J_\mu (x)$ is some {\it gauge--invariant} (observable) 
four--current and if the physical spectrum is free of massless particles, 
then $(q_m - q_n)^\mu \langle m\vert J_\mu (0) \vert n \rangle$ must tend 
to zero as $q_m - q_n \to 0$.
However, $K_\mu (x)$ is {\it gauge--noninvariant} and
$\langle m\vert K_\mu (0) \vert n \rangle$ can contain massless 
singularities of the form:
\be
\langle m\vert K_\mu (0) \vert n \rangle 
\mathop{\simeq}_{q_m \to q_n}
k_{mn} {(q_m - q_n)_\mu \over (q_m - q_n)^2}
~~~~ (k_{mn} = {\rm constant}) ~.
\ee
This behaviour is known in the literature as a ``Kogut--Susskind (KS) 
pole'', since it was first conjectured by Kogut and Susskind in Ref.
\cite{Kogut-Susskind75} (for the matrix element
$\langle \pi\pi \vert K_\mu (0) \vert \eta \rangle$) 
in the context of a mechanism for explaining the 
large (unsuppressed) decay amplitude of $\eta \to 3\pi$.
If there were no KS poles, from Eqs. (3.17) we would be allowed to conclude 
that $\chi(\beta)$ is real and equal to the corresponding Euclidean quantity 
$\chi_E(\beta)$.
Instead, this conclusion can be invalidated by the possible existence of
KS poles.

Let us go back to Eqs. (3.17) and try to explore more carefully the 
relationship between $\chi(\beta)$ and $\chi_E(\beta)$. Since
\be
\displaystyle\lim_{q_m^0 \to q_n^0} 
{1 - e^{\beta (q_m^0 - q_n^0)} \over q_m^0 - q_n^0} = -\beta
\ee
is a finite quantity, the last term in Eq. (3.16) vanishes assuming that the 
(possible) KS poles lie within a region of continuous distribution of 
energy eigenvalues $(q_m^0,q_n^0)$, in which case the set $q_m^0 = q_n^0$ 
has null measure in the plane $(q_m^0,q_n^0)$, and also assuming that the 
first term in Eq. (3.16) is finite (which seems to be confirmed {\it a 
posteriori} by lattice measurements).
We are then left with
\be
\tilde{G}_\beta (p=0) = {(2\pi)^3 \over Z(\beta)} 
\displaystyle\sum_{q_m^0 \ne q_n^0}
{e^{-\beta q_m^0} - e^{-\beta q_n^0} \over q_m^0 - q_n^0} 
|\langle m | Q(0) | n \rangle|^2 \delta^{(3)} (\vec{q}_m - \vec{q}_n) ~.
\ee
That is, comparing with Eq. (3.12):
\be
\tilde{G}_\beta (p=0) = {\rm Re} [\tilde{I}_\beta (p=0)] ~.
\ee
On the other hand, from Eq. (3.10) we also have that:
\ba
{\rm Re} [\tilde{I}_\beta (p)] &=& \displaystyle\int_0^{\infty} d(E^2)
{\bar \rho}_\beta (E^2-\vec{p}^2,E) (1-e^{-\beta E}) 
{\rm P} {1 \over p_0^2 - E^2} ~; \nonumber \\
{\rm Im} [\tilde{I}_\beta (p)] &=& 
-\pi (1-e^{-\beta |p_0|}) {\bar \rho}_\beta (p^2,|p_0|) 
-2\pi \rho_{\beta,0} (p^2,p_0) ~.
\ea
Making the limits $p^0 \to 0$ and $\vec{p} \to \vec{0}$ in these equations, 
we get:
\be
\displaystyle\lim_{p^0 \to 0}\lim_{\vec{p} \to \vec{0}} 
{\rm Re} [\tilde{I}_\beta (p)] =
-\displaystyle\int_0^{\infty} d(E^2)
{\bar \rho}_\beta (E^2,E) (1-e^{-\beta E}) {\rm P} {1 \over E^2}
= {\rm Re} [\tilde{I}_\beta (p=0)] ~.
\ee
And also, using Eqs. (3.9):
\be
\displaystyle\lim_{p^0 \to 0}\lim_{\vec{p} \to \vec{0}} 
{\rm Im} [\tilde{I}_\beta (p)] = 0 ~,
\ee
which in general may be different from
\be
{\rm Im} [\tilde{I}_\beta (p=0)] = -2\pi \rho_{\beta,0} (0,0) ~.
\ee
This means that, while the real part of $\tilde{I}_\beta(p)$ is continuous 
at $p=0$, the imaginary part may (in general) have a discontinuity at
$p^0 = 0$, due to the existence of KS poles. In other words,
$\tilde{I}_\beta(p)$ may be discontinuous at $p^0 = 0$.
From Eqs. (3.32) and (3.33) one has that:
\be
\displaystyle\lim_{p^0 \to 0}\lim_{\vec{p} \to \vec{0}} 
\tilde{I}_\beta (p) = {\rm Re} [\tilde{I}_\beta (p=0)] ~.
\ee
On the contrary, from Eqs. (3.15) and (3.16) one deduces that 
$\tilde{G}_\beta (p)$ is continuous at $p=0$:
$\displaystyle\lim_{p^0 \to 0}\lim_{\vec{p} \to \vec{0}} 
\tilde{G}_\beta (p) = \tilde{G}_\beta (p=0)$.

\noindent
Comparing Eq. (3.35) with Eq. (3.30), one concludes that:
\be
\tilde{G}_\beta (p=0) =
\displaystyle\lim_{p^0 \to 0}\lim_{\vec{p} \to \vec{0}} 
\tilde{I}_\beta (p) ~.
\ee
That is, by the definitions (3.21) and (3.4):
\be
\chi_E(\beta) =
\displaystyle\lim_{p^0 \to 0}\lim_{\vec{p} \to \vec{0}} 
\left( -i \displaystyle\int d^4 x e^{ipx} 
\langle {\rm T} Q(x) Q(0) \rangle_\beta \right) \equiv \chi^{(reg)} (\beta)
~.
\ee
In conclusion, the Euclidean topological susceptibility at finite 
temperature $\chi_E(\beta)$ is not, in a strict sense, equal to the Minkowskian 
topological susceptibility $\chi(\beta)$, but it is equal to the limit for
$p^0 \to 0$ and for $\vec{p} \to \vec{0}$ of the Fourier transform of
$-i \langle {\rm T} Q(x) Q(0) \rangle_\beta$: we have called this quantity
$\chi^{(reg)} (\beta)$, meaning that it can be seen as a re--definition of
the Minkowskian topological susceptibility by introducing a proper 
infrared regularization.
Of course, in the zero--temperature limit $T \to 0$ (i.e., $\beta \to 
\infty$) one recovers the result $\chi_E = \chi$, obtained at the end of 
the previous Section. In fact, in this limit the discontinuity (3.34) 
disappears:
\be
\displaystyle\lim_{\beta \to \infty} \rho_{\beta,0} (0,0) = 0 ~,
\ee
since $\langle 0| Q(0) |0 \rangle = 0$, so that 
$\chi = \chi(\beta \to \infty)$ becomes real and equal to 
$\chi_E = \chi_E (\beta \to \infty) = \chi^{(reg)} (\beta \to \infty)$, 
in agreement with Eq. (2.20).
Instead, at finite temperature ($T \ne 0$), Eq. (3.37) is not trivial and
furnishes the correct equality relation between the Euclidean topological
susceptibility and the ``infrared--regularized'' Minkowskian topological
susceptibility.

We want to observe, also, that the ``infrared--regularized'' topological
susceptibility $\chi^{(reg)}$, if considered in the ``quenched'' limit
$N_c \to \infty$ ($N_c$ being the number of colours), is exactly the
quantity which plays a fundamental role in the Witten--Veneziano
mechanism for the solution of the $U(1)$ problem: it provides the $\eta^\prime$
meson with a large ``gluonic'' mass \cite{Witten79,Veneziano79}.
In fact, in the approach of Witten and Veneziano one first sums all
diagrams in $\tilde{I}(p)$ of a given order in $1/N_c$ and then considers
the limit $p \to 0$. In other words, one first writes $\tilde{I}(p)$ as
$\tilde{I}(p) = \tilde{I}_0 (p) + \tilde{I}_1 (p) + \tilde{I}_2 (p) 
+ \ldots$, where $\tilde{I}_n (p)$ is the sum of all diagrams with $n$
quark loops (each quark loop is suppressed by a factor of $1/N_c$).
The limit $p \to 0$ is taken at the end.
The Witten--Veneziano mechanism was originally formulated for the $T=0$
case \cite{Witten79,Veneziano79}. However, it can be generalized to the
$T \ne 0$ case, as was done in Ref. \cite{Meggiolaro94}.
We hope to return to a more detailed discussion about this point in the
near future. 

\bigskip
\noindent {\bf Acknowledgements}
\smallskip
 
I would like to thank Prof. Adriano Di Giacomo for his useful suggestions
and comments and, first of all, for having prompted my interest in this 
subject. I also thank him for the critical reading of this paper.

\vfill\eject

\renewcommand{\thesection}{A}
\renewcommand{\thesubsection}{A.\arabic{subsection}}
 
\pagebreak[3]
\setcounter{section}{1}
\setcounter{equation}{0}
\setcounter{subsection}{0}
\setcounter{footnote}{0}
 
\begin{flushleft}
{\bf Appendix: The contact term.}
\end{flushleft}
 
\noindent
Some subtleties arise in connection with the proper definition of the 
topological susceptibility $\chi$ and the use of the T--ordered product
in Eqs. (1.1), (1.3), (2.1) and (3.1). The problem is that the two--point
Green function $\langle {\rm T} Q(x) Q(0) \rangle$ is not well defined
as $x \to 0$: one must give a prescription for treating the product of the
two (composite) operators $Q(x)$ and $Q(0)$ when approaching the same 
space--time point $x \to 0$. Actually, this ambiguity is eliminated in the 
``correct'' definition of the topological susceptibility, which can be found 
in Ref. \cite{Witten79}:
\be
\chi \equiv {1 \over VT}{i \over Z[\theta]}
{{\rm d}^2 Z[\theta] \over {\rm d}\theta^2} |_{\theta = 0} ~.
\ee
Here $VT$ is an infinite four--volume which must be factorized [in a 
symbolic notation: $VT = \int d^4x = \int d^4x e^{ipx}|_{p=0} =
(2\pi)^4 \delta^{(4)}(0)$] and $Z[\theta]$ is the partition function of the
theory (in the path--integral formalism) with the addition of a 
$\theta$--term to the usual action:
\be
Z[\theta] \equiv \displaystyle\int [dA][d\psi][d{\bar \psi}]
e^{i(S + \theta q[A])} ~.
\ee
$S$ is the usual action for the full theory and $q[A] \equiv \int d^4x Q(x)$
is the (total) topological charge. As shown in the Appendix of Ref.
\cite{Witten79}, a certain equal--time commutator term (also called
``contact term'') must be added to the right--hand side of Eq. (2.1) to
get the correct formula (A.1) for the topological susceptibility:
\be
\chi = -i\int d^4x \langle {\rm T} Q(x) Q(0) \rangle + \chi^{(1P)} ~,
\ee
where the contact term $\chi^{(1P)}$ is given by (in the temporal gauge
$A^0 = 0$):
\be
\chi^{(1P)} = 8 \left( {g^2 \over 16 \pi^2} \right)^2
\langle {\rm Tr} [\vec{B}^2] \rangle ~,
\ee
and $B_i^a = -{1 \over 2} \varepsilon_{ijk} F_{jk}^a$ is the non--Abelian
magnetic field. [Here and in the following we shall adopt the suffix ``$1P$''
(which stands for ``{\it one--point}'') to indicate the contact term, while
the first term at the right--hand side of Eq. (A.3) will be denoted as
$\chi^{(2P)}$, since the {\it two--point} function of $Q(x)$ appears in the
integral.]

An alternative definition of $\chi$ can be found in Ref. \cite{Crewther79},
where the T--ordering ambiguity is eliminated using the following 
prescription:
\be
\chi \equiv -i \int d^4(x-y) \partial^\mu_x \partial^\nu_y 
\langle {\rm T} K_\mu (x) K_\nu (y) \rangle ~.
\ee
Here $K_\mu (x)$ is the usual four--current defined in Eq. (3.25), whose 
divergence gives the topological charge density, $Q(x) = \partial^\mu
K_\mu (x)$. Eq. (A.5) can also be put in the following equivalent form:
\be
\chi \equiv -i \int d^4x \partial^\mu_x \langle {\rm T} K_\mu (x)
Q(0) \rangle ~.
\ee
Both (A.5) and (A.6) give rise to the following (more explicit) expression
for $\chi$ [actually, the two integrand expressions in (A.5) and (A.6)
differ by a quasi--local operator $\Delta (x)$, which anyway vanishes after 
space--time integration]:
\be
\chi = -i \int d^4x \langle {\rm T} Q(x) Q(0) \rangle -i \int d^3 \vec{x}
\langle [K^0 (x), Q(0)]_{x^0 = 0} \rangle ~.
\ee
The equal--time commutator term at the right--hand side of Eq. (A.7) is just
the contact term, which is needed to make the definition of the topological
susceptibility unambiguous. In fact, any T--ordered product ambiguity in
(A.6) is necessarily a derivative $\partial_x$ of $\delta^{(4)} (x-y)$: but
such terms are annihilated after integration in $\int d^4x$.
The quantity $\chi$, defined in (A.5) $\div$ (A.7), enters into the anomalous
Ward identities of QCD, which were derived in Ref. \cite{Crewther79}.

Using the same procedure outlined in the Appendix of Ref. \cite{Witten79}
(canonical quantization in the temporal gauge $A^0 = 0$), one
can easily verify that the contact term in Eq. (A.7) is exactly equal to the
contact term (A.4), derived by Witten from the definition (A.1) of $\chi$:
\be
-i \int d^3 \vec{x} \langle [K^0 (x), Q(0)]_{x^0 = 0} \rangle = \chi^{(1P)}
= 8 \left( {g^2 \over 16 \pi^2} \right)^2 \langle {\rm Tr} [\vec{B}^2] \rangle
~.
\ee
The same expression is also derived for the finite--temperature contact term
$\chi^{(1P)}_\beta$, except that one must now intend $\langle \ldots \rangle$
not as the simple vacuum expectation value $\langle 0| \ldots |0 \rangle$, 
but as the usual quantum thermal average over the Gibbs ensemble
$\langle \ldots \rangle_\beta$, defined in Eqs. (3.2) and (3.3).
Therefore one has that:
\be
\chi^{(1P)}_\beta \equiv -i \int d^3 \vec{x} \langle [K^0 (x), Q(0)]_{x^0 = 0} 
\rangle_\beta = 8 \left( {g^2 \over 16 \pi^2} \right)^2 \langle {\rm Tr} 
[\vec{B}^2] \rangle_\beta ~.
\ee
The reason for this simple result is that in the explicit evaluation of the
integral in Eq. (A.9) one simply uses canonical (equal--time) commutation 
relations,
which are fundamental and do not depend on the temperature of the system.
For these reasons, all the results obtained in this Appendix  are valid both
at zero temperature ($T=0$, in which case: $\langle \ldots \rangle =
\langle 0| \ldots |0 \rangle$) and at finite temperature ($T \ne 0$, in
which case: $\langle \ldots \rangle = \langle \ldots \rangle_\beta$).
We observe, in addition, that at $T=0$ the contact term can put in the
Lorentz-- and gauge--invariant form $\chi^{(1P)} = (g^2/64\pi^2) G_2$,
where $G_2 \equiv (\alpha_s/\pi)\langle F^a_{\mu\nu} F^{a,\mu\nu} \rangle$
(with $\alpha_s \equiv g^2/4\pi$) is the so--called {\it gluon condensate}:
one uses the relation $F^a_{\mu\nu} F^{a,\mu\nu} = 2 \Tr [F_{\mu\nu} 
F^{\mu\nu}] = 4 \Tr [\vec{B}^2 - \vec{E}^2]$ and the fact that, at zero 
temperature
($T=0$), the electric part of the gluon condensate $G^{(el)}_2 \equiv
-(g^2/\pi^2) \langle \Tr [\vec{E}^2] \rangle$ is equal to the magnetic part
of the gluon condensate $G^{(mag)}_2 \equiv (g^2/\pi^2) \langle \Tr [\vec{B}^2] 
\rangle$, because of Lorentz-- and parity--invariance.
Instead, at finite temperature ($T \ne 0$), Lorentz invariance is broken down
to $O(3)$ rotational invariance: therefore $G^{(mag)} \ne G^{(el)}$ and one
must keep the expression (A.9) for the contact term, i.e.,
$\chi^{(1P)} = (g^2/32\pi^2) G^{mag}_2$.

All the above refers to the physical, i.e., Minkowskian, space--time. Let us
see, now, what happens in the Euclidean four--space. The Euclidean topological
susceptibility $\chi_E$ is defined by continuing the definition (A.1) or,
equivalently, (A.5) $\div$ (A.7), to the Euclidean world, by using the
correspondence relationships (2.18) and (2.19) from Minkowskian to
Euclidean theory:
\be
\chi_E \equiv -{1 \over V_E T_E}{1 \over Z_E [\theta]}
{{\rm d}^2 Z_E [\theta] \over {\rm d} \theta^2} |_{\theta = 0} ~.
\ee
$Z_E [\theta]$ is the Euclidean partition function (in the path--integral
formalism) with the addition of a $\theta$--term to the usual 
Euclidean action:
\be
Z_E [\theta] \equiv \displaystyle\int [dA_E][d\psi_E][d{\bar \psi}_E]
e^{-S_E + i\theta q_E[A_E]} ~.
\ee
$S_E$ is the usual Euclidean action for the full theory and $q_E[A_E] \equiv 
\int d^4x_E Q_E(x_E)$ is the (total) Euclidean topological charge.
Eq. (A.10) provides us with a rigorous definition of the Euclidean Green
function $\langle Q_E (x_E) Q_E (y_E) \rangle_E$, which appears in the 
expression:
\be
\chi_E \equiv {1 \over V_E T_E} \int d^4 y_E \int d^4 x_E
\langle Q_E (x_E) Q_E (y_E) \rangle_E 
= \int d^4 x_E \langle Q_E (x_E) Q_E (0) \rangle_E ~,
\ee
in terms of an Euclidean path--integral:
\be
\langle Q_E (x_E) Q_E (0) \rangle_E \equiv {1 \over Z_E}
\displaystyle\int [dA_E][d\psi_E][d{\bar \psi}_E] Q_E (x_E) Q_E (0)
e^{-S_E} ~.
\ee
In other words, the correct expression for the Euclidean Green function
$\langle Q_E (x_E) Q_E (y_E) \rangle_E$, in the sense given by Eqs. (A.10)
$\div$ (A.13), is not simply obtained by the analitic continuation of
$-\langle {\rm T} Q(x) Q(y) \rangle$, as assumed in Eqs. (2.16) and (3.20):
rigorously speaking, Eqs. (2.16) and (3.20) are only true when $\tau \ne 0$. 
In the general case, one must also include the analytic continuation of the
contact term, since, by virtue of Eqs. (A.10) $\div$ (A.13) [to be compared
with Eqs. (A.1), (A.2) and (A.5), (A.6)], $\langle Q_E (x_E) Q_E (y_E) 
\rangle_E$ is the  analytic continuation of $-\partial^\mu_x \partial^\nu_y 
\langle {\rm T} K_\mu (x) K_\nu (y) \rangle$; that is:
\be
\langle Q_E (\vec{x},\tau) Q_E (\vec{y},\sigma) \rangle_E =
-\partial^\mu_x \partial^\nu_y \langle {\rm T} K_\mu (x) K_\nu (y) \rangle 
|_{(x^0,y^0) \to (-i\tau,-i\sigma)} ~.
\ee
Explicitly, this means that:
\ba
\lefteqn{
\langle Q_E (\vec{x},\tau) Q_E (\vec{0},0) \rangle_E = }\nonumber \\
& & = -\langle {\rm T} Q(\vec{x},-i\tau) Q(\vec{0},0) \rangle
-i \delta(\tau) \langle [K^0 (\vec{x},-i\tau), Q(\vec{0},0)]_{\tau = 0} \rangle
+ \Delta(\vec{x},-i\tau) ~,
\ea
where $\Delta (x)$ is a quasi--local operator which vanishes after
space--time integration [see Eq. (A.7)]. Therefore, after integration in 
$\int d^3 \vec{x} \int d\tau$, we obtain:
\be
\chi_E \equiv \int d^3 \vec{x} \int d\tau 
\langle Q_E (\vec{x},\tau) Q_E (\vec{0},0) \rangle_E
= \chi^{(2P)}_E + \chi^{(1P)}_E ~,
\ee
where
\be
\chi^{(2P)}_E \equiv 
-\int d^3 \vec{x} \int d\tau 
\langle {\rm T} Q (\vec{x},-i\tau) Q (\vec{0},0) \rangle
\ee
is the contribution to $\chi_E$ coming from the {\it two--point} function
\be
\langle Q_E (\vec{x},\tau) Q_E (\vec{0},0) \rangle^{(2P)}_E \equiv
-\langle {\rm T} Q (\vec{x},-i\tau) Q (\vec{0},0) \rangle ~,
\ee
while
\be
\chi^{(1P)}_E = \chi^{(1P)} = -i \int d^3 \vec{x} \langle 
[K^0 (\vec{x},0), Q(\vec{0},0)] \rangle
\ee
is the contribution to $\chi_E$ from the contact (equal--time commutator)
term. We observe that $\chi^{(1P)}_E$ turns out to be trivially equal to the
Minkowskian contact term $\chi^{(1P)}$. This is due to the fact that
$\chi^{(1P)}$ is an equal-time commutator term, so that the continuation
to imaginary times $x^0 \to -i\tau$ is trivial in this case.
Let us observe also that, by virtue of the correspondence relationships
(2.18) and (2.19) from Minkowskian to Euclidean theory, we have:
\be
K^0 (x) \to K_{E4} (x_E) ~~~ , ~~~ K^i (x) \to iK_{Ei} (x_E) ~,
\ee
so that $Q(x) \to iQ_E (x_E)$, with $Q_E = \partial_{E\alpha} K_{E\alpha}$.
Therefore, we can re--write Eq. (A.14) in the following way:
\be
\langle Q_E (x_E) Q_E (y_E) \rangle_E =
{\partial \over \partial x_{E\alpha}}
{\partial \over \partial y_{E\beta}}
\langle {\rm T} K_{E\alpha} (x_E) K_{E\beta} (y_E) \rangle_E ~,
\ee
in agreement with (A.13). In other words:
\ba
\lefteqn{
\chi^{(1P)}_E = \int d^3 \vec{x} \langle 
[K_{E4} (\vec{x},\tau), Q_E (\vec{0},0)]_{\tau = 0} \rangle_E }\nonumber \\
& & = -i \int d^3 \vec{x} \langle 
[K^0 (\vec{x},-i\tau), Q(\vec{0},0)]_{\tau = 0} \rangle \nonumber \\
& & = -i \int d^3 \vec{x} \langle 
[K^0 (\vec{x},0), Q(\vec{0},0)] \rangle = \chi^{(1P)} ~.
\ea
In conclusion, the contact term for the topological susceptibility in the 
Euclidean world is exactly (and trivially) equal to the contact term in the 
Minkowski world. This is true both at zero temperature and at finite
temperature. Therefore, the inclusion of the contact (equal--time commutator)
term does not affect the results derived in the text (where we proved that
$\chi^{(2P)} = \chi^{(2P)}_E$, with the notation introduced above), except
that it solves some questions about positivity arising with some formulas.
As a rule, in all formulas where $\chi$ is mentioned in the text, one must
intend that it is actually $\chi^{(2P)}$: then, to obtain  the proper value
of $\chi$, consistently with the definitions (A.1) [(A.5), (A.6)] and (A.10),
one must add the contact term $\chi^{(1P)}$, in the way explained in this
Appendix.

\vfill\eject

{\renewcommand{\Large}{\normalsize}
}

\vfill\eject

\noindent
\begin{center}
{\bf FIGURE CAPTIONS}
\end{center}
\vskip 0.5 cm
\begin{itemize}
\item [\bf Fig.~1.] The integration path for $\langle Q(\vec{x},\tau) 
Q(\vec{0}, 0) \rangle_\beta $ in the complex $\tau$--plane.
\end{itemize}

\vfill\eject

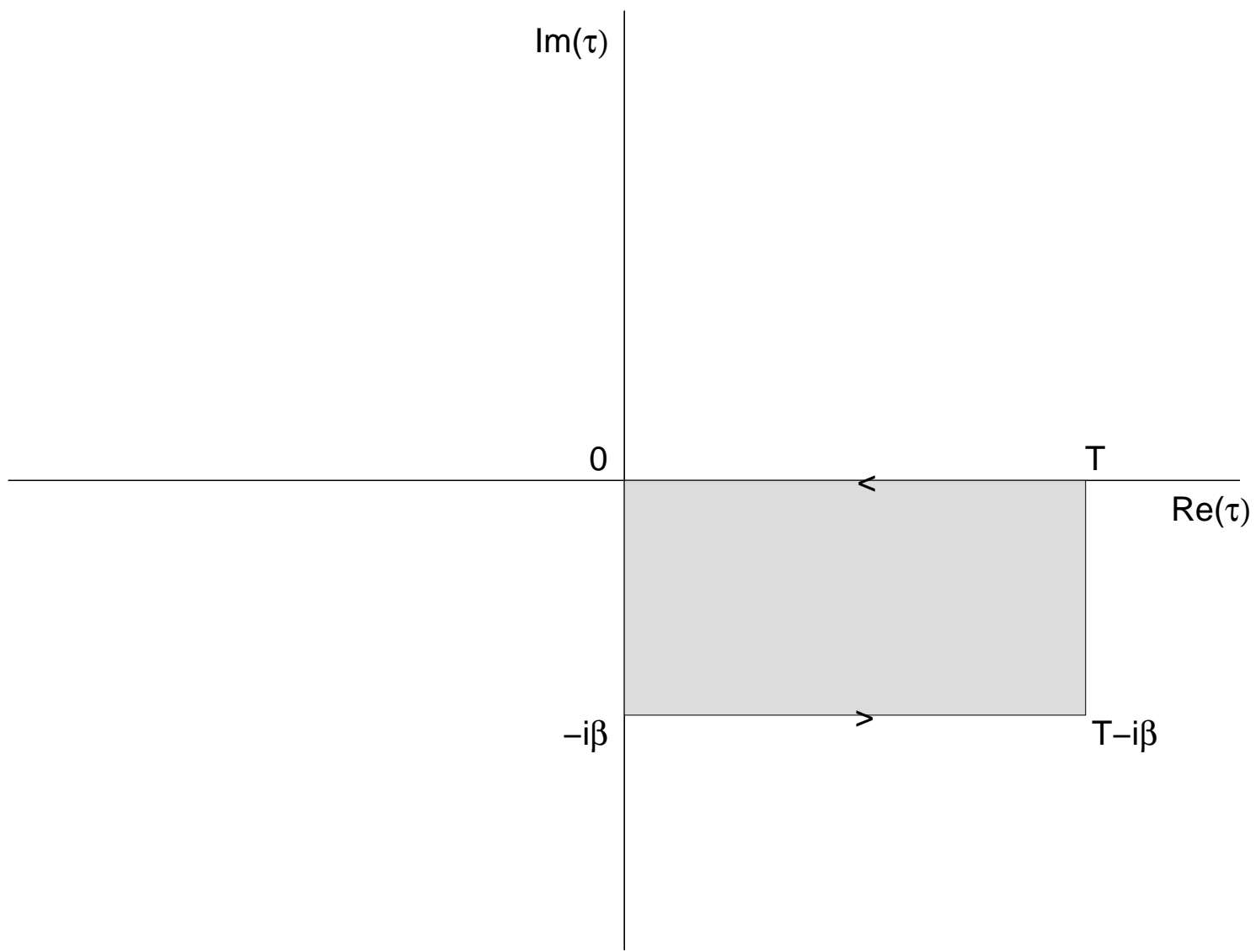
\end{document}